\documentclass[longauth, traditabstract]{aa} 
\usepackage{graphicx}
\usepackage{txfonts}
\begin{document}

\title{Detection of OH$^+$ and H$_2$O$^+$ towards Orion~KL}

\author{
H.~Gupta,\inst{1}
P.~Rimmer,\inst{2}
J.~C.~Pearson,\inst{1}
S.~Yu,\inst{1}
E.~Herbst,\inst{2}
N.~Harada,\inst{2}
E.~A.~Bergin,\inst{3}
D.~A.~Neufeld,\inst{4}
G.~J.~Melnick,\inst{5}
R.~Bachiller,\inst{6}
W.~Baechtold,\inst{7}
T.~A.~Bell,\inst{8}
G.~A.~Blake,\inst{8}
E.~Caux,\inst{9,10}
C.~Ceccarelli,\inst{11}
J.~Cernicharo,\inst{12}
G.~Chattopadhyay,\inst{1}
C.~Comito,\inst{13}
S.~Cabrit,\inst{14}
N.~R.~Crockett,\inst{3}
F.~Daniel,\inst{12,15}
E.~Falgarone,\inst{15}
M.~C.~Diez-Gonzalez,\inst{6}
M.-L.~Dubernet,\inst{16,17}
N.~Erickson,\inst{18}
M.~Emprechtinger,\inst{8}
P.~Encrenaz,\inst{15}
M.~Gerin,\inst{15}
J.~J.~Gill,\inst{1}
T.~F.~Giesen,\inst{19}
J.~R.~Goicoechea,\inst{12}
P.~F.~Goldsmith,\inst{1}
C.~Joblin,\inst{9,10}
D.~Johnstone,\inst{21}
W.~D.~Langer,\inst{1}
B.~Larsson,\inst{20}
W.~B.~Latter\inst{22} 
R.~H.~Lin,\inst{1}
D.~C.~Lis,\inst{8}
R.~Liseau,\inst{23}
S.~D.~Lord,\inst{22}
F.~W.~Maiwald,\inst{1}
S.~Maret,\inst{11}
P.~G.~Martin,\inst{24}
J.~Martin-Pintado,\inst{12}
K.~M.~Menten,\inst{13}
P.~Morris,\inst{22}
H.~S.~P. M\"uller,\inst{19}
J.~A.~Murphy,\inst{25}
L.~H.~Nordh,\inst{20}
M.~Olberg,\inst{23}
V.~Ossenkopf,\inst{19,26}
L.~Pagani,\inst{14}
M.~P\'erault,\inst{15}
T.~G.~Phillips,\inst{8}
R.~Plume,\inst{27}
S.-L.~Qin,\inst{19}
M.~Salez,\inst{14}
L.~A.~Samoska,\inst{1}
P.~Schilke,\inst{13,19}
E.~Schlecht,\inst{1}
S.~Schlemmer,\inst{19}
R.~Szczerba,\inst{27}
J.~Stutzki,\inst{19}
N.~Trappe,\inst{25}
F.~F.~S.~van der Tak,\inst{26}
C.~Vastel,\inst{9,10}
S.~Wang,\inst{3}
H.~W.~Yorke,\inst{1}
J.~Zmuidzinas,\inst{8}
A.~Boogert,\inst{8}
R.~G\"usten,\inst{17}
P.~Hartogh,\inst{28}
N.~Honingh,\inst{21}
A.~Karpov,\inst{8}
J.~Kooi,\inst{8}
J.-M.~Krieg,\inst{12}
R.~Schieder\inst{19}
\and
P.~Zaal\inst{26}
}
\institute {Jet Propulsion Laboratory, Caltech, Pasadena, CA 91109, USA \\
                 \email{hgupta@jpl.nasa.gov}
\and Departments of Physics, Astronomy, and Chemistry, Ohio State University, Columbus, OH 43210, USA
\and Department of Astronomy, University of Michigan, 500 Church Street, Ann Arbor, MI 48109, USA 
\and  Department of Physics and Astronomy, Johns Hopkins University, 3400 North Charles Street, Baltimore, MD 21218, USA
\and Harvard-Smithsonian Center for Astrophysics, 60 Garden Street, Cambridge MA 02138, USA
\and Observatorio Astron\'omico Nacional (IGN), Centro Astron\'omico de Yebes, Apartado 148. 19080 Guadalajara,  Spain
\and Microwave Laboratory, ETH Zurich, 8092 Zurich, Switzerland
\and California Institute of Technology, Cahill Center for Astronomy and Astrophysics 301-17, Pasadena, CA 91125 USA    
\and Centre d'\'etude Spatiale des Rayonnements, Universit\'e de Toulouse [UPS], 31062 Toulouse Cedex 9, France
\and CNRS/INSU, UMR 5187, 9 avenue du Colonel Roche, 31028 Toulouse Cedex 4, France
\and Laboratoire d'Astrophysique de l'Observatoire de Grenoble, BP 53, 38041 Grenoble, Cedex 9, France.
\and Centro de Astrobiolog\'ia (CSIC/INTA), Laboratiorio de Astrof\'isica Molecular, Ctra. de Torrej\'on a Ajalvir, km 4
28850, Torrej\'on de Ardoz, Madrid, Spain
\and Max-Planck-Institut f\"ur Radioastronomie, Auf dem H\"ugel 69, 53121 Bonn, Germany 
\and LERMA \& UMR8112 du CNRS, Observatoire de Paris, 61, Av. de l'Observatoire, 75014 Paris, France
\and LERMA, CNRS UMR8112, Observatoire de Paris and \'Ecole Normale Sup\'erieure, 24 Rue Lhomond, 75231 Paris Cedex 05, France
\and LPMAA, UMR7092, Universit\'e Pierre et Marie Curie,  Paris, France
\and  LUTH, UMR8102, Observatoire de Paris, Meudon, France
\and University of Massachusetts, Astronomy Dept., 710 N. Pleasant St., LGRT-619E, Amherst, MA 01003-9305  U.S.A
\and I. Physikalisches Institut, Universit\"at zu K\"oln, Z\"ulpicher Str. 77, 50937 K\"oln, Germany
\and Department of Astronomy, Stockholm University, SE-106 91 Stockholm, Sweden
\and National Research Council Canada, Herzberg Institute of Astrophysics, 5071 West Saanich Road, Victoria, BC V9E 2E7, Canada 
\and Infrared Processing and Analysis Center, California Institute of Technology, MS 100-22, Pasadena, CA 91125
\and Chalmers University of Technology, SE-412 96 Gšteborg, Sweden, Sweden; Department of Astronomy, Stockholm University, SE-106 91 Stockholm, Sweden
\and Canadian Institute for Theoretical Astrophysics, University of Toronto, 60 St George St, Toronto, ON M5S 3H8, Canada
\and  National University of Ireland. Maynooth, Ireland
\and SRON Netherlands Institute for Space Research, PO Box 800, 9700 AV, Groningen, The Netherlands
\and N. Copernicus Astronomical Center, Rabianska 8, 87-100, Torun, Poland
\and Department of Physics and Astronomy, University of Calgary, 2500 University Drive NW, Calgary, AB T2N 1N4, Canada
\and MPI f\"ur Sonnensystemforschung, D 37191 Katlenburg-Lindau, Germany
}


\abstract{
We report observations of the reactive molecular ions OH$^+$, H$_2$O$^+$, and H$_3$O$^+$ 
towards Orion~KL with Herschel/HIFI.  All three $N=1-0$ fine-structure transitions of OH$^+$ at 909, 971, and 1033~GHz and 
both fine-structure components of the doublet {\it ortho}-H$_2$O$^+$ $1_{11}-0_{00}$ transition at 1115 and 1139~GHz were detected; 
an upper limit was obtained for H$_3$O$^+$. OH$^+$ and H$_2$O$^+$ are observed purely in absorption, showing a narrow 
component at the source velocity of 9 km~s$^{-1}$, and a broad blueshifted absorption similar to that reported recently for HF and 
{\it para}-H$_{2}^{18}$O, and attributed to the low velocity outflow of Orion~KL. 
We estimate column densities of OH$^+$ and H$_2$O$^+$ for the 9 km s$^{-1}$ component of $9 \pm 3 \times 10^{12}$~cm$^{-2}$ 
and $7 \pm 2 \times 10^{12}$~cm$^{-2}$, and those in the outflow of $1.9 \pm 0.7 \times 10^{13}$~cm$^{-2}$ and 
$1.0 \pm 0.3 \times 10^{13}$~cm$^{-2}$. Upper limits of $2.4\times 10^{12}$~cm$^{-2}$  and $8.7\times 10^{12}$~cm$^{-2}$ 
were derived for the column densities of {\it ortho} and {\it para}-H$_3$O$^+$ from transitions near 985 and 1657~GHz.  The 
column densities of the three ions are up to an order of magnitude lower than those obtained from recent observations of W31C 
and W49N.  The comparatively low column densities may be explained by a higher gas density despite the assumption of a 
very high ionization rate.
}

   \titlerunning{OH$^+$ and H$_2$O$^+$ in Orion KL}
	\authorrunning{Gupta et al.}
   \maketitle

\section{Introduction}

The Heterodyne Instrument for Far Infrared (HIFI) on the {\em Herschel Space Observatory} \footnote{{\em Herschel} is an ESA space
observatory with science instruments provided by European-led Principal Investigator consortia and with important participation
from NASA.} provides a unique opportunity to fully assess the first steps of the oxygen chemistry in a wide variety of sources.  
Initial HIFI observations quickly detected widespread absorption by OH$^+$ and H$_{2}$O$^{+}$ toward the star forming 
regions DR21, W31C, and W49N (Ossenkopf et al. 2010; Gerin et al. 2010;  Neufeld et al. 2010, this issue). 
Prior to the HIFI observations, OH$^+$ had only been detected in absorption 
toward Sgr B2(M) (Wyrowski et al. 2010). Similarly, previous observations of H$_{2}$O$^{+}$ were limited to its detection  in 
comet tails (e.g., Herzberg \& Lew 1974; Wehinger et al. 1974), demonstrating the importance of photoionization in 
producing this ion in the absence of H$_2$.  And until recently, only upper limits had been reported on the column density 
of H$_2$O$^+$ in the diffuse interstellar gas (Smith, Schempp, \& Federman 1984).

By contrast, the recent HIFI detections of OH$^+$ and H$_{2}$O$^{+}$ in warm diffuse gas with a fairly small fraction 
of molecular hydrogen, elucidated the role of O$^+$ in initiating the oxygen-hydrogen chemistry.  This chemistry 
is thought to begin with the production of H$^+$ and H$_{3}^+$ {\it via} cosmic ray or X-ray ionization of hydrogen, followed by 
charge transfer to produce O$^+$. Rapid hydrogen abstraction reactions of O$^+$ with H$_2$ 
then yield OH$^+$ and H$_{2}$O$^{+}$, and terminate with the production of H$_{3}$O$^{+}$.  In diffuse molecular clouds, which have 
high electron abundances, the H$_3$O$^+$ is destroyed {\it via} dissociative recombination to yield OH and H$_2$O. In dense 
molecular clouds, both the ionization fraction and the atomic hydrogen abundance are comparatively lower, and the 
sequence of reactions, expected to start at H$_{3}^+$ and OH$^+$, yields a larger abundance of H$_{3}$O$^{+}$.  
This picture is probably overly simplistic for molecular clouds such as Orion~KL, which are composed of both diffuse and dense gas.

Orion~KL is the brightest infrared region in the Orion-Monoceros molecular cloud complex located less than 500 pc from 
the sun (Menten et al. 2007).  In the foreground of Orion~KL is the Orion Nebula, an HII region known to contain a cluster 
of thousands of young stars which produce a substantial flux of X-ray photons (Getman et al. 2005).  Molecular line studies reveal three 
main regions in Orion~KL: i. a core of very dense and hot gas ($n \sim 10^{7}$ cm$^{-2}$, $T \sim 200$ K); ii. cool, quiescent gas 
between systemic velocities of 8 km~s$^{-1}$ and 10 km~s$^{-1}$, surrounded by high-velocity outflows ($\ge 100$ km~s$^{-1}$); 
and iii. a highly inhomogeneous and turbulent outflow source containing both high-velocity ($\ge 30$ km~s$^{-1}$)  and 
low-velocity ($\sim 18$ km~s$^{-1}$) gas (Blake et al. 1987; Genzel \& Stutzki 1989; O'Dell et al. 2008).


In this {\it Letter} we report the detection of absorption lines of OH$^+$ and H$_{2}$O$^{+}$, and an upper limit on the 
column density of H$_{3}$O$^{+}$ toward Orion~KL.  In addition to molecular absorption at a systemic velocity of 9 km~s$^{-1}$, 
these observations find broad blueshifted absorption by OH$^+$ and H$_{2}$O$^{+}$ extending to large negative velocities.  
This is consistent with previously observed lines of H$_2$O with ISO (Lerate et al. 2006), as well as those of HF and 
{\it para}-H$_{2}^{18}$O detected recently with HIFI, and attributed to the low-velocity molecular outflow (Phillips et al. 2010).

\section{Observations and data reduction}

The observations were done in March 2010 as part of the 
Key Program {\it Herschel/HIFI Observations of Extraordinary sources: The Orion 
and Sagittarius Star-forming Regions} (HEXOS). The dual beam switch (DBS) observing mode 
was used, with the DBS reference beams lying approximately $3^\prime$ east and 
west of the Orion KL position $\alpha_{J2000} = 5^h35^m14.3^s$ and $\delta_{J2000} =
-5^{\circ}22'33.7''$. Spectra were taken with 
the Wide Band Spectrometer (WBS) with a Nyquist-limited frequency resolution of approximately 
1.1~MHz over a 4~GHz wide IF band; the HIFI beams in bands 4, 5, and 6 have half-power beam 
widths of $21''$, $19''$, and $13''$ and main beam efficiencies of $0.670$, $0.662$, and $0.645$ (HIFI Observers' Manual, v 2.0). 
The spectra were reduced through the standard Herschel Pipeline to Level 2 using HIPE version 2.4 (Ott 2010). 
The double sideband (DSB) spectra so obtained were then 
deconvolved (Comito \& Schilke 2002) to single sideband (SSB) spectra using the {\it doDeconvolution} 
task in HIPE. The SSB spectra were converted to the FITS format and analyzed with the CLASS90 package. 
Although two orthogonal polarizations were observed simultaneously, only spectra from the H
polarization in bands 4a and 6b and the V polarization in band 5a are shown, because of the smaller 
standing waves in these polarizations. 


\section{Spectroscopy}

The spectroscopy of OH$^+$, H$_2$O$^+$, and H$_3$O$^+$ has been discussed in detail in the 
recent detection papers (Ossenkopf et al. 2010; Gerin et al. 2010). Here we summarize the essential 
aspects of the rotational spectra of these ions.  The OH$^+$ ion has a $^{3}\Sigma^{-}$ electronic  
ground state, the two unpaired electron spins ($S=1$) yielding three components of the $N=1-0$ transition. 
The nuclear spin of the hydrogen atom ($I_{H}=1/2$) further splits each component into hyperfine components.  
The H$_2$O$^+$ ion has $C_{2v}$ symmetry and a $^{2}B_{1}$ ground state which results in the lowest level 
having {\it ortho} symmetry.  The spin of the unpaired electron ($S=1/2$) results in two fine-structure components, 
each exhibiting a complex hyperfine pattern due to the spins of the two equivalent hydrogen nuclei ($I_{H}=1/2$). 
Rotational spectroscopy of H$_2$O$^+$ is limited to two laser magnetic resonance (LMR) studies 
(Strahan et al. 1986; M{\"u}rtz et al. 1998).   Here, we adopt the values of M{\"u}rtz et al., which 
we and others have checked independently to be accurate to about 2~km~s$^{-1}$ in equivalent radial velocity 
(see Neufeld et al. 2010 and Schilke et al. 2010, this volume).  H$_3$O$^+$ is a closed-shell symmetric 
top molecule with a large amplitude inversion near 1.65 THz, resulting in a spectrum similar to 
NH$_3$ with transitions between symmetric and antisymmetric inversion states (Yu et al. 2009).  
Table~1 lists the observed transitions of the three ions, along with their line strengths and 
spontaneous emission rates. 

\begin{table}[h]
\tiny
\label{tab:spec}
\caption{Spectroscopic parameters of the observed transitions.}
\begin{tabular}{rcrrrrr}
\hline\hline\rule[-3mm]{0mm}{6mm}
$Transition$ & $Frequency$ &  $E_l$            & $g_l$ & $g_u$ &$\mu^{2}S$~\tablefootmark{a} & $10^{2}A_{ij}$  \\
                       &  ( MHz)            &  (cm$^{-1}$)  &            &              & ($\rm{D}^2$)                &     (s$^{-1}$)     \\
\hline
\multicolumn{7}{c}{\rule[-3mm]{0mm}{6mm} OH$^+$ $N=1-0$~\tablefootmark{b} } \\
\multicolumn{7}{c}{\rule[-3mm]{0mm}{4mm} $J=1-0$} \\
$F=1/2-1/2$         &  $909045.2\pm1.5$   &  0.004  & 4 & 6 & 1.20  & 1.05 \\
    $ 1/2-3/2$         &  $909158.8\pm1.5$   &  0 & 2 & 4 &   2.40   & 0.52 \\
\multicolumn{7}{c}{\rule[-3mm]{0mm}{6mm} $J=2-1$} \\
$F=5/2-3/2$         &  $971803.8\pm1.5$   &  0  & 4 & 6 & 10.24  & 1.82 \\
    $ 3/2-1/2$         &  $971805.3\pm1.5$   &  0.004 & 2 & 4 &   5.69   & 1.52 \\
    $ 3/2-3/2$         &  $971919.2\pm1.0$   &  0 & 4 & 4 &   1.14   & 0.30 \\
\hline
\multicolumn{7}{c}{\rule[-3mm]{0mm}{6mm} o-H$_2$O$^+$  $1_{11}-0_{00}$~\tablefootmark{c}} \\
\multicolumn{7}{c}{\rule[-3mm]{0mm}{4mm} $J=3/2-1/2$} \\
$F=3/2-1/2$       & $1115150.0\pm1.8$    &  0.004 & 2 & 4 &   4.14  & 1.67 \\
     $1/2-1/2$       & $1115186.0\pm1.8$    &  0.004 & 2 & 2 &   3.32  & 2.68 \\
     $5/2-3/2$       & $1115204.0\pm1.8$    &  0 & 4 & 6 & 11.23  & 3.02 \\
     $3/2-3/2$       & $1115263.0\pm1.8$   &  0 & 4 & 4 &   3.35  & 1.35 \\
     $1/2-3/2$       & $1115298.7\pm1.8$   &  0 & 2 & 2 &   0.42  & 0.34 \\
\multicolumn{7}{c}{\rule[-3mm]{0mm}{6mm} $J=1/2-1/2$} \\
$F=3/2-1/2$       & $1139541.1\pm1.8$    &  0.004 & 2 & 2 &   0.42  & 3.61 \\
     $1/2-1/2$       & $1139560.6\pm1.8$    &  0.004 & 2 & 4 &   3.35  & 1.44 \\
     $5/2-3/2$       & $1139653.5\pm1.8$    &  0 & 4 & 2 &   3.32  & 2.86 \\
     $3/2-3/2$       & $1139673.3\pm1.8$   &  0 & 4 & 4 &    4.14  & 1.78 \\    
\hline
\multicolumn{7}{c}{\rule[-3mm]{0mm}{6mm}H$_3$O$^+$~\tablefootmark{d} } \\
$0_{0}^{-} - 1_{0}^{+}$  &   $984711.9\pm0.1$\tablefootmark{e}  &    5.1 &   4 & 12 & 8.30 & 2.30 \\
$1_{1}^{-} - 1_{1}^{+}$ & $1655834.8\pm0.3$\tablefootmark{f}  &    0 &   6 &   6 & 6.22 & 5.48 \\
$2_{2}^{-} - 2_{2}^{+}$  & $1657248.4\pm0.3$\tablefootmark{e} & 29.6  & 10 & 10 & 4.67 & 7.32 \\
\hline
\end{tabular}
\tablefoot{
\tablefoottext{a}{Dipole moments ($\mu$): 2.256~D (OH$^+$; Werner, Rosmus, \& Reinsch 1983); 
2.37~D (H$_2$O$^+$; Wu et al. 2004); 1.44~D (H$_3$O$^+$; Botschwina, Rosmus, \& Reinsch 1985)}. 
Frequencies from: \tablefoottext{b}{M{\"u}ller et al. (2005)}; \tablefoottext{c}{M{\"u}rtz et al. (1998)}; \tablefoottext{d}{Yu et al. (2009)} \\
\tablefoottext{e}{$\int T_A d{\rm v} < 0.482$ K km s$^{-1}$ for the 984.7 GHz line, and $< 2.412$ K~km~s$^{-1}$ for the 1657.2 GHz line.}\\
\tablefoottext{f}{Blended with a strong $2_{12}-1_{10}$ {\it ortho}-H$_2^{18}$O line at 1655831~MHz.}
}
\end{table}

\section{Results}

\begin{figure}
  \resizebox{\hsize}{!}{\includegraphics[angle=270]{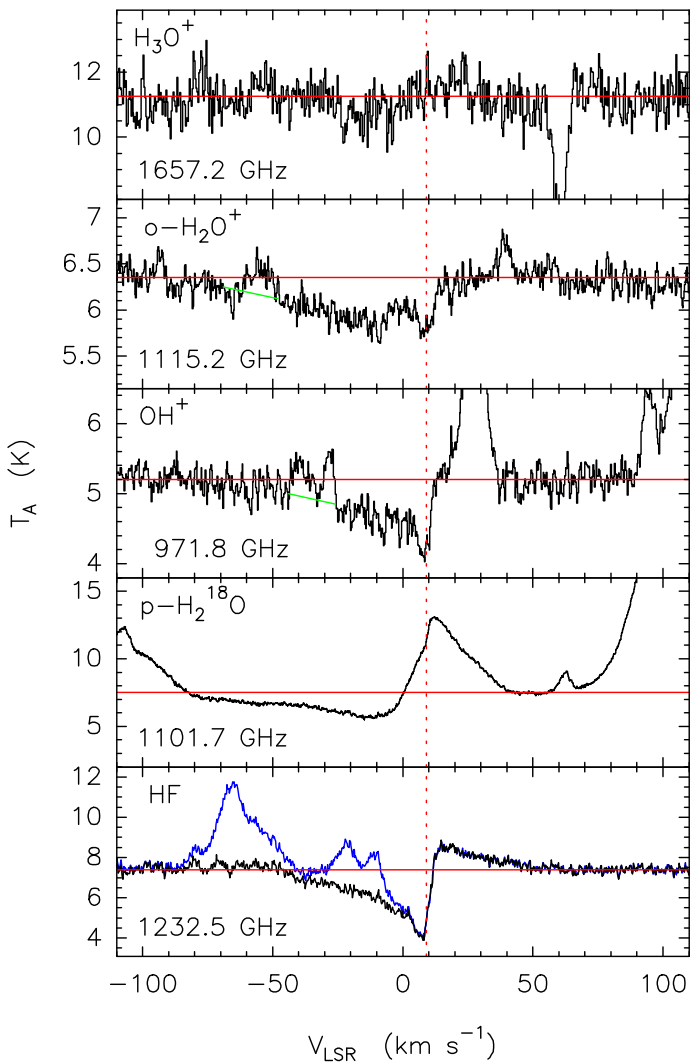}}
 \caption{Lines of OH$^+$, {\it ortho}-H$_2$O$^+$, and H$_3$O$^+$ in Orion~KL, compared with 
  those of HF ($J=1-0$) and {\it para}-H$_{2}^{18}$O.
  The dashed vertical red line is at the systemic velocity of 9~km~s$^{-1}$, and 
  the solid horizontal red lines indicate the continuum level in each spectrum.  Solid green lines indicate the 
  channels over which the interpolation was done. The HF spectrum, adapted  from 
  Phillips et al (2010), shows a broad absorption (black histogram) after modeling and removal of 
  contaminating lines of CH$_3$OH and SO$_2$ (blue histogram). The contaminants in the 971.8~GHz 
  spectrum are CH$_3$OH ($v_t=0$) $26_{2}^{-}-25_{3}^{-}$ (A) and ($v_t=1$) $22_{6}-22_{5}$ (E) at 
  $-28$~km~s$^{-1}$ and $-40$~km~s$^{-1}$; that in the 1115.2~GHz spectrum is SO$_2$ 
  $19_{5,15}-19_{2,18}$ at $-55$~km~s$^{-1}$.  The 1657.2~GHz spectrum shows absorption 
  by CH ($N=2-1$) at $60$~km~s$^{-1}$.}
    \label{fig1}
\end{figure}

Figure~\ref{fig1} shows the absorption lines of OH$^+$ and H$_2$O$^+$ toward Orion~KL, as well as 
lines of HF and {\it para}-H$_{2}^{18}$O for comparison.  The strongest hyperfine components 
of OH$^+$ and H$_2$O$^+$ appear at the source velocity of 9 km s$^{-1}$, which matches well 
that of the HF line in Orion~KL.  Additionally, lines of both ions show broad blue absorption wings 
extending to about $-75$ km s$^{-1}$, more extended than the HF absorption, but comparable 
to that of {\it para}-H$_{2}^{18}$O ($\sim -80$ km~s$^{-1}$).  We attribute the extended absorption 
of the ions to originate mainly from the low velocity molecular outflow. We failed to detect any 
emission or absorption from H$_3$O$^+$, and discuss the non-detection in $\S$~\ref{sec:disc}.

The high density of molecular lines in Orion~KL makes contamination 
by unrelated lines a common problem.  The absorption lines detected here are blended with weak to 
moderately strong emission lines of abundant ``weeds'', including CH$_3$OH and SO$_2$.  Efforts 
are underway to model and remove the emission from the contaminants by 
a method similar to that of Phillips et al. (2010); in the interim, the following approach was taken.

To better gauge the absorption, the contaminants were masked and intensities interpolated across 
the masked channels (Fig.~\ref{fig1}).  The velocity-integrated optical depths of the ionic lines were obtained by normalizing 
the SSB spectra with the continuum and integrating over the velocity ranges for the 
source and the outflow, the interpolation yielding errors of $20\%-30\%$. 
On the assumptions that the absorption covers the source completely, and the molecules are in the lower 
state, the total column density ($N$) was then derived using the expression:

\begin{equation}
\int \tau d{\rm v}~({\rm km~s^{-1}}) = \frac{A_{ul}g_u \lambda^3}{8\pi g_l} N,
\end{equation}

\noindent where $A_{ul}$ is the spontaneous emission rate, $g_u$ and 
$g_l$ are the upper and lower state degeneracies, and $\lambda$ is the transition wavelength. 

We estimate column densities of OH$^+$ and H$_2$O$^+$ at 9 km s$^{-1}$  of $9 \pm 3 \times 10^{12}$~cm$^{-2}$ 
and $7 \pm 2\times 10^{12}$~cm$^{-2}$, and those in the outflow of $1.9 \pm 0.7\times 10^{13}$~cm$^{-2}$ and 
$1.0 \pm 0.3\times 10^{13}$~cm$^{-2}$.  The column densities of OH$^+$ are more than an order of magnitude lower, 
and those of H$_2$O$^+$ are $2-6$ times lower than toward W31C and 
W49N (Gerin et al. 2010; Neufeld et al. 2010, this issue). From the least congested spectra 
of H$_3$O$^+$ at 984.7 and 1657.2~GHz (see Table~1), and an 
assumed excitation temperature of 100~K, we derive $3\sigma$ upper limits of $2.4 \times 10^{12}$~cm$^{-2}$ 
and $8.7 \times 10^{12}$~cm$^{-2}$ for the column density of {\it ortho} and {\it para}-H$_3$O$^+$, 
nearly an order of magnitude lower than in W31C (Gerin et al. 2010).

The abundance ratios of the three ions in Orion~KL can be compared to the same ratios observed in W31C and 
W49N.  The OH$^+$/H$_2$O$^+$ ratio is found to be $1.3 \pm 0.6$ in the 
source and $1.8 \pm 0.8$ in the outflow.  This ratio is $2-15$ times lower than that measured toward W31C and W49N. 
The lower limit of 1.4 for the H$_2$O$^+$/H$_3$O$^+$ ratio, however, is nearly 2 times larger than  
in W31C.

\section{Discussion}
\label{sec:disc}

The column densities of OH$^+$, H$_2$O$^+$, and H$_3$O$^+$ in Orion~KL differ markedly from 
those in the diffuse gas toward W31C and W49N.  In contrast with W31C and W49N, OH$^+$ and H$_2$O$^+$ 
are significantly more abundant relative to H$_3$O$^+$, for which we are only able to obtain an upper limit. 
The absolute column densities of OH$^+$ and H$_2$O$^+$ are also lower compared with W31C and W49N. 
A likely explanation for the low column densities of the three ions is that they are present in fairly dense material, both in 
the quiescent gas and the outflow.  Unlike the quiescent gas, the Orion~KL outflow is
exposed to a strong ionizing flux from the foreground HII region; the enhanced ionization flux enhances the formation 
of ions, but the resultant large fractional ionization leads to a fast and efficient removal of molecular ions
by dissociative recombination with electrons.

The observed velocity profiles of OH$^+$ and H$_2$O$^+$ in Orion~KL  support the above conclusion. 
As Fig.~\ref{fig1} shows,  the OH$^+$ and H$_2$O$^+$ absorption tracks the HF absorption to velocities of about 
$-45$~km~s$^{-1}$.  This absorption also seems to follow closely, to about $-80$~km~s$^{-1}$, the {\it para}-H$_{2}^{18}$O 
absorption in the outflow, suggesting that like HF and  {\it para}-H$_{2}^{18}$O, OH$^+$ and H$_2$O$^+$ probably exist mainly 
in the low velocity outflow (Phillips et al. 2010).  In fact, the molecular outflow accounts for over half of the 
observed column density of OH$^+$ and H$_2$O$^+$.

The conditions required to explain our observations may be more extreme than one might suppose.  First, the 
molecular ions probably reside in gas of lower density ($n \le 10^5$~cm$^{-3}$) than that necessary to thermally excite the 
observed transitions---these have high spontaneous emission rates ($ > 10^{-2}$~s$^{-1}$; Table~1), and hence large 
critical densities ($10^{7}-10^{9}$~cm$^{-3}$) .  This is supported by the observation that OH$^+$ and H$_2$O$^+$ are 
seen {\it only} in absorption. Second, the temperatures in the outflow gas are probably high.

We consider two scenarios in which the ions may be formed in the low velocity outflow.
In the first, a large radiation flux impinges directly on the Orion~KL outflow, which contains large water 
abundances (Melnick et al. 2010). The far UV flux that illuminates this gas can have values 
approaching $4\times 10^4$ times the average interstellar radiation field (Walmsley et al. 2000; Young Owl et al. 2000). 
In addition, the central region of the Orion Nebula has numerous sources of energetic X-ray photons 
(Getman et al. 2005; Preibisch et al. 2005), which can contribute to the surface ionization of this photon-dominated 
region (PDR).  We estimate that at A$_V=1$ into the PDR, the ionization rate $\zeta_{X} \sim 3 \times 10^{-15}$ s$^{-1}$.
\footnote{The surface brightness of the central region (dominated by $\theta ^1C$) is estimated to be $3\times10^{34}$ ergs s$^{-1}$pc$^{-2}$ (Feigelson et al. 2005).  On the assumption that the molecular cloud lies 0.1 pc from this 
cluster, the expressions of Maloney et al. (1996)  yield an X-ray ionization rate 
of  about 2.8 $\times$ 10$^{-16}$N$_{22}^{-1}$ (where N$_{22}$ is the hydrogen 
column density in units of 10$^{22}$ cm$^{-2}$).  Thus at A$_V = 1$, 
$ \zeta_{X} \sim 3 \times 10^{-15}$ s$^{-1}$.} 
Under these conditions, water can undergo photoionization to form H$_2$O$^+$ directly, enhancing the abundance of this species.

In the second scenario, the outflow penetrates the extended foreground HII region.  
The abundant H$^+$ can now undergo charge exchange with H$_2$O to yield H$_2$O$^+$. 
In either scenario, the high electron density probably results in a net reduction in the abundances 
of molecular ions, consistent with the observations: low column densities of OH$^+$ and 
H$_2$O$^+$, and the upper limit for H$_3$O$^+$.

We have attempted to model the first scenario using the Meudon PDR code (Le Petit et al. 2006). 
However, the model suffers difficulties while reproducing the observed column densities 
of the three ions.  First, it requires a relatively low gas density ($n \sim 10^3$~cm$^{-3}$) in regions where 
OH$^+$ and H$_2$O$^+$ are produced, as larger assumed densities yield too much H$_3$O$^+$.  
Second, it requires a very large ionization rate ($\zeta > 1- 2\times10^{-14}$ s$^{-1}$) to maintain a ratio 
of atomic to molecular hydrogen near unity; otherwise, too much H$_{3}$O$^{+}$ is once again produced. 
The two parameters are nearly an order of magnitude different from others inferred from previous observations: $n \ge 10^4$~cm$^{-3}$ 
and $\zeta < 10^{-14}$~s$^{-1}$ (Genzel \& Stutzki 1989; Lerate et al. 2008;  Muench et al. 2008 and references therein).  
Nevertheless, a recent study on molecular hydrogen rotational excitation in the Orion bar infers a cosmic ray 
ionization rate of $7 \times 10^{-14}$ s$^{-1}$ (Shaw et al. 2009). 
The same study also invokes warm gas temperatures of 400-700 K; the lower value is contained in our model 
for the edge of the PDR.  A critical evaluation of our model awaits further work and a thorough exploration of 
the parameter space, and will be presented in a future paper. 

We are unable to confirm previous tentative detections of H$_3$O$^+$ toward Orion~KL
(Hollis et al. 1986; Wootten et al. 1986; Wootten et al. 1991; Phillips et al. 1992; 
Timmermann et al. 1996; Lerate et al. 2006).  Of these, Phillips et al. (1992) present the best 
evidence: 3 emission lines at 307, 364, and 396~GHz, lying 45, 85, and 105 cm$^{-1}$ 
above ground; but they do not rule out the possibility of blends with other lines. 
The lines we observed are at lower energies (see Table~1), and are expected 
to be as strong or stronger than those observed by Phillips et al. (1992). The upper limits derived here 
for {\it ortho} and {\it para}-H$_3$O$^+$ are more than an order of magnitude lower than the column 
densities reported by Phillips et al. (1992).  Timmermann et al. (1996) reported detection of the 
$4_{3}^{-}-3_{3}^{+}$ line near $70~\mu\rm{m}$ with the Kuiper Airborne Observatory, but the velocity 
of the line differs by more than $-60$~km~s$^{-1}$ from predicted values.  Lerate et al. (2006) detected 
the $2_{1}^{-}-1_{1}^{+}$, $2_{0}^{-}-1_{0}^{+}$, and $1_{1}^{-}-1_{1}^{+}$ lines with ISO: the first, near 
2.98~THz is $80$~km~s$^{-1}$ higher than the systemic velocity of 9~km~s$^{-1}$; the second, near 
2.97~THz, is 1~km~s$^{-1}$ higher than the frequencies predicted by Yu et al. (2009); and the third, 
at 1655835~MHz, covered by our observations, is obscured by a strong $2_{12}-1_{10}$ {\it ortho}-H$_2^{18}$O line at 1655831~MHz.

\section{Conclusions}

Our observations toward Orion~KL have found OH$^+$ and H$_2$O$^+$ aborption
at the quiescent 9 km~s$^{-1}$ component and extended absorption  
in the low velocity molecular outflow associated with this source. This is, to our knowledge, the first 
detection of these ions toward a source with a large fraction of molecular gas. 
Given the complex and inhomogeneous nature of Orion~KL, however, there are probably 
regions where the densities are sufficiently low and the excitation conditions optimal 
for these reactive ions to exist at detectable levels. Another possibility is that 
depletion of some of the gas-phase species onto the grains can result in lower abundances 
of water, leading to small column densities of OH$^{+}$ and H$_{2}$O$^{+}$.
A surprising observation---and one remarkably different from that toward W31C---is the non-detection of  H$_3$O$^+$.    
In our model of the outflow, we attribute this mainly to a
very high ionization rate, which produces an almost equal abundance of
atomic and molecular hydrogen at the assumed density.

\begin{acknowledgements}
HIFI has been designed and built by a consortium of institutes and university departments from across 
Europe, Canada and the United States under the leadership of SRON Netherlands Institute for Space
Research, Groningen, The Netherlands and with major contributions from Germany, France and the US. 
Consortium members are: Canada: CSA, U.Waterloo; France: CESR, LAB, LERMA,  IRAM; Germany: 
KOSMA, MPIfR, MPS; Ireland, NUI Maynooth; Italy: ASI, IFSI-INAF, Osservatorio Astrofisico di Arcetri- 
INAF; Netherlands: SRON, TUD; Poland: CAMK, CBK; Spain: Observatorio Astron—mico Nacional (IGN), 
Centro de Astrobiolog'a (CSIC-INTA). Sweden:  Chalmers University of Technology - MC2, RSS \& GARD; 
Onsala Space Observatory; Swedish National Space Board, Stockholm University - Stockholm Observatory; 
Switzerland: ETH Zurich, FHNW; USA: Caltech, JPL, NHSC.
Support for this work was provided by NASA through an award issued by JPL/Caltech. A part of the work 
described in this paper was done at the Jet Propulsion Laboratory, California Institute of Technology, under 
contract with the National Aeronautics and Space Administration. Copyright 2010$\copyright$ California Institute of Technology. All rights reserved.
\end{acknowledgements}

\end{document}